\documentclass[runningheads]{llncs}
\usepackage[dvipsnames]{xcolor}
\usepackage{amsmath}
\usepackage{caption}
\usepackage{subcaption}
\usepackage{graphicx}
\usepackage{amssymb}
\usepackage{pdfpages}
\usepackage{hyperref}

\usepackage{multirow}

\begin{document}
%

\title{Hybrid graph convolutional neural networks 
\\for landmark-based anatomical segmentation}
%
%
\author{Nicolás Gaggion, 
Lucas Mansilla, 
Diego Milone,
Enzo Ferrante
}
\authorrunning{N. Gaggion et al., 2021}
\titlerunning{Hybrid GCNNs for landmark-based anatomical segmentation}

%
\institute{Research Institute for Signals, Systems and Computational Intelligence, sinc(i)\\ CONICET, Universidad Nacional del Litoral, Santa Fe, Argentina}
%

\maketitle              
\begin{abstract}
In this work we address the problem of landmark-based segmentation for anatomical structures. We propose HybridGNet, an encoder~-~decoder neural architecture which combines standard convolutions for image feature encoding, with graph convolutional neural networks to decode plausible representations of anatomical structures. We benchmark the proposed architecture considering other standard landmark and pixel-based models for anatomical segmentation in chest x-ray images, and found that HybridGNet is more robust to image occlusions. We also show that it can be used to construct landmark-based segmentations from pixel level annotations. Our experimental results suggest that HybridGNet produces accurate and anatomically plausible landmark-based segmentations, by naturally incorporating shape constraints within the decoding process via spectral convolutions.
\end{abstract}
\section{Introduction}
Deep learning models based on convolutional neural networks have become the state-of-the-art for anatomical segmentation of biomedical images. 
The current practise is to employ standard convolutional neural networks (CNNs) trained to minimize a pixel level loss function, where dense segmentation masks are used as ground truth. Casting image segmentation as a pixel labeling problem is desirable in scenarios like lesion segmentation, where topology and location do not tend to be preserved across individuals. However, organs and anatomical structures usually present a characteristic topology which tends to be regular. Differently from dense segmentation masks, statistical shape models \cite{heimann2009statistical} and graph-based representations \cite{boussaid2014discriminative} provide a natural way to incorporate topological constraints by construction. Moreover, such shape representations make it easier to establish landmark correspondences among individuals, particularly important in the context of statistical shape analysis. 

Since the early 1990's, variations of point distribution models (PDMs) have been proposed \cite{cootes1992training} to segment anatomical structures using landmarks. PDMs are flexible shape templates describing how the relative locations of important points can vary. Techniques based on PDMs, like active shape models (ASM) \cite{cootes1992training,sozou1997non} and active appearance models (AAM) \cite{cootes1998active} became the defacto standard to deal with anatomical segmentation at the end of the century. Subsequently, during the next decade, the development of more powerful and robust image registration algorithms \cite{zitova2003image} positioned deformable template matching algorithms as the choice of option for anatomical segmentation and atlas construction \cite{frangi2001automatic,heitz2004automatic,paulsen2002building}. More recently, with the advent of deep fully convolutional networks \cite{shakeri2016sub,ronneberger2015u}, great efforts were made to incorporate anatomical constraints into such models \cite{jurdia2020high,larrazabal2019anatomical,oktay2017anatomically}. The richness of the image features learned by standard CNNs allowed them to achieve highly accurate results.
However, most of these methods work directly on the pixel space, producing acceptable dense segmentations masks but without landmark annotations and connectivity structure. On the contrary, structured models like graphs appear as a natural way to represent landmarks, contours and surfaces. By defining the landmark position as a function on the graph nodes, and encoding the anatomical structure through its adjacency matrix, we can easily constrain the space of solutions and ensure topological correctness.

During the last years, the emerging field of geometric deep learning \cite{bronstein2017geometric} extended the success of convolutional neural networks to non-Euclidean domains like graphs and meshes. While classical CNNs have been particularly successful when dealing with signals such as images or speech, more recent developments like spectral convolutions \cite{bruna2013spectral,defferrard2016convolutional} and neural message passing \cite{gilmer2017neural} enabled the use of deep learning on graphs. Recently, graph generative models were proposed \cite{kipf2016variational,ranjan2018generating}. Of particular interest for this work is the convolutional mesh autoencoder proposed in \cite{ranjan2018generating}. The authors construct an encoder-decoder network using spectral graph convolutions, and train it in a variational setting using face meshes. By sampling the latent space, they are able to generate new expressive faces never seen during training. Inspired by this idea, we propose to exploit the generative power of convolutional graph autoencoders \cite{ranjan2018generating,foti2020intraoperative} to decode plausible anatomical segmentations from images.

\noindent \textbf{Contributions.} In this work, we revisit landmark-based segmentation in light of the latest developments on deep learning for Euclidean and non-Euclidean data. We aim at leveraging the best of both worlds, combining standard convolutions for image feature encoding, with generative models based on graph convolutional neural networks (GCNN) to decode plausible representations of anatomical structures. Under the hyphothesis that encoding connectivity information through the graph adjacency matrix will result in richer representations than standard landmark-based PDMs, we also compare our results with other statistical point distribution models which do not make explicit use of the graph connectivity. Our contributions are 4-fold: (1) we propose HybridGNet, an encoder-decoder architecture which combines standard convolutions with GCNNs to extract graph representations directly from images; (2) we showcase the proposed architecture in the context of anatomical landmark-based segmentation of chest x-ray images and benchmark the results considering other landmark and pixel-based segmentation models; (3) we propose to use HybridGNet to create graph-based representations from dense anatomical masks without paired images and (4) we show that HybridGNet is more robust to image oclussions than state-of-the-art pixel-level segmentation methods.
\begin{figure}[t!]
\includegraphics[width=\textwidth]{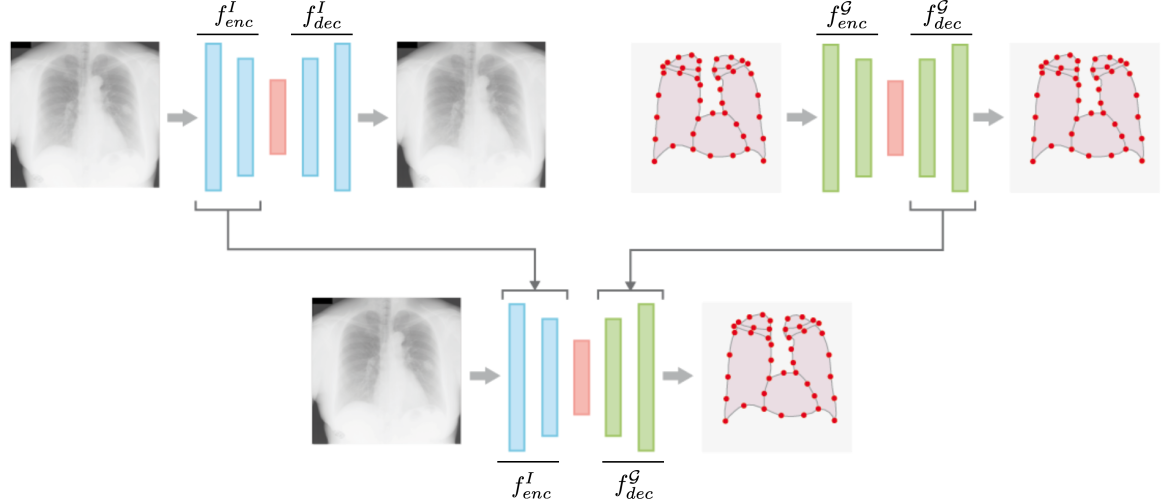}
\caption{The proposed HybridGNet (bottom) is an encoder-decoder architecture which combines standard convolutions for image feature encoding (blue), with graph spectral convolutions (green) to decode plausible anatomical graph-based representations.} 
\label{fig:workflow}
\end{figure}

\section{Hybrid graph convolutional neural networks}

\noindent \textbf{Problem setting.} Let us have a dataset  $\mathcal{D} = \{(I, \mathcal{G})_{k}\}_{0 < k < N}$, composed of $N$ dimensional images $I$ and their corresponding landmark-based segmentation represented as a graph $\mathcal{G}=<V,\mathbf{A},\mathbf{X}>$. $V$ is the set of nodes (landmarks), $\mathbf{A} \in \{0,1\}^{|V| \times |V|}$ is the adjacency matrix indicating the connectivity between pairs of nodes ($\mathbf{A}_{ij} = 1$ indicates an edge connecting vertices $i$ and $j$, and $\mathbf{A}_{ij} = 0$ otherwise) and $\mathbf{X} \in \mathbb{R}^{|V| \times d}$ is a function (represented as a matrix) assigning a feature vector to every node. In our case, it assigns a d-dimensional spatial coordinate to every landmark. Without loss of generality, in this work we will showcase the proposed framework in 2D images (thus $\mathbf{X} \in \mathbb{R}^{|V| \times 2}$). However, note that extending our method to 3D images is straightforward, the only difference being that images $I$ will be volumes and graphs $\mathcal{G}$, meshes. 

In the context of landmark-based segmentation and point distribution models, it is common to have manual annotations with fixed number of points. Therefore, for all graphs, we assume that the set of nodes (landmarks) $V$ and the connectivity matrices $\mathbf{A}$ are the same. The only difference among them is given by the spatial coordinates defined in $\mathbf{X}$. This assumption enables the use of spectral graph convolutions \cite{defferrard2016convolutional,ranjan2018generating} to learn latent representations of anatomy.\\ 

\noindent \textbf{Spectral graph convolutions.}
We have previously mentioned that extending discretized convolutions to graph structures is not straightforward. A myriad of graph convolution formulations have been recently proposed \cite{bronstein2017geometric,gilmer2017neural}. Here we adopt the localized spectral version since it has shown to be effective at learning powerful shape representations from graph structures with a fixed number of nodes \cite{ranjan2018generating}. Spectral convolutions are build using the eigendecomposition of the graph Laplacian matrix $\mathbf{L}$, exploiting the property that convolutions in the node domain are equivalent to multiplications in the graph spectral domain \cite{shuman2013emerging}. 

The graph Laplacian is defined as $\mathbf{L} = \mathbf{D} - \mathbf{A}$, where $\mathbf{D}$ is the diagonal degree matrix with $\mathbf{D}_{ii} = \sum_{j} \mathbf{A}_{ij}$, and $\mathbf{A}$ is the adjacency matrix. The Laplacian $\mathbf{L}$ can be decomposed as $\mathbf{L} = \mathbf{U \Lambda U}^T$, where $\mathbf{U} \in \mathbb{R}^{|V| x |V|} = [u_0, u_1, ..., u_{|V|-1}]$ is the matrix of eigenvectors (Fourier basis) and $\mathbf{\Lambda} = \mathrm{diag}(\lambda_0, \lambda_1 ... \lambda_{|V|-1})$ the diagonal matrix of eigenvalues (frequencies of the graph). By analogy with the classical Fourier transform for continuous or discrete signals, the graph Fourier transform of a function $\mathbf{X}$ defined on the graph domain is $\mathbf{\hat{X}} = \mathbf{U}^T \mathbf{X}$, while its inverse is given by $\mathbf{X} = \mathbf{U} \mathbf{\hat{X}}$. Based on this formulation, the spectral convolution between a signal $\mathbf{X}$ and a filter $\mathbf{g}_\phi = \mathrm{diag}(\phi)$ is defined as $\mathbf{g}_\phi * \mathbf{X} = \mathbf{g}_\phi (\mathbf{L}) \mathbf{X} = \mathbf{g}_\phi (\mathbf{U \Lambda U}^T) \mathbf{X} = \mathbf{U} \mathbf{g}_\phi (\mathbf{\Lambda}) \mathbf{U}^T \mathbf{X}$, where $\phi \in \mathbb{R}^n$ is a vector of coefficients parameterizing the filter. We follow the work of Defferrard et al \cite{defferrard2016convolutional} and restrict the class of filters to polynomial filters $\mathbf{g}_\phi = \sum_{k=0}^K \phi_k \mathbf{\Lambda}^k$. Polynomial filters are strictly localized in the vertex domain (a K-order polynomial filter considers
K-hop neighborhoods around the node) and reduce the computational complexity of the convolutional operator. Such filters can be well approximated by a truncated expansion in terms of Chebyshev polynomials computed recursively. Following  \cite{defferrard2016convolutional,ranjan2018generating} we adopt this approximation to implement the spectral convolutions. Note that a spectral convolutional layer will take feature matrices $\mathbf{X^l}$ as input, and produce filtered versions $\mathbf{X^{l+1}}$ akin to what standard convolutions do with images and feature maps.\\

\noindent \textbf{Auto-encoding shape and appearance.}
Autoencoders are neural networks designed to reconstruct their input. They follow an encoder-decoder scheme, where an encoder $z = f_{enc}(x)$ maps the input $x$ to a lower dimensional latent code $z$, which is then processed by a decoder $f_{dec}(z)$ to reconstruct the original input. 
The bottleneck imposed by the low-dimensionality of the encoding $z$ forces the model to retain useful information, learning powerful representations of the data distribution. The model is trained to minimize a reconstruction loss $\mathcal{L}_{rec}(x, f_{dec}(f_{enc}(x)))$ between the input and the output reconstruction. 
To constrain the distribution of the latent space $z$, we add a variational loss term to the objective function, resulting in a variational autoencoder (VAE) \cite{kingma2013auto}. 
We assume that the latent codes $z$ are sampled from a distribution $Q(z)$ for which we will impose a unit multivariate Gaussian prior. In practise, during training, this results in the latent codes $z$ being sampled from a distribution $\mathcal{N}(\mathbf{\mu}, \mathbf{\sigma})$ via the reparametrization trick \cite{kingma2013auto}, where $\mathbf{\mu}$, $\mathbf{\sigma}$ are deterministic parameters generated by the encoder $f_{enc}(x)$. Given a sample $z$, we can generate (reconstruct) the corresponding data point by using the decoder $f_{dec}(x)$. This model is trained by minimizing a loss function defined as:
\begin{equation}
    \mathcal{L}_{vae} = \mathcal{L}_{rec}(x, f_{dec}(z)) + w \textrm{KL}(\mathcal{N}(0,1) || Q(z | x)),
\end{equation}
where the first term is the reconstruction loss, and the second term imposes a unit Gaussian prior $\mathcal{N}(0,1)$ via the KL divergence loss.
Depending on the type of data $x$ that will be processed, here we will implement $f_{enc}$ and $f_{dec}$ using standard convolutional layers (to encode image appearence) or spectral convolutions (to encode graph structures). \\

\noindent \textbf{Hybrid graph convolutional neural networks (HybridGNet).}
To construct the HybridGNet, we pre-train two independent VAEs. The first one, defined by the corresponding encoder $f^I_{enc}$ and decoder $f^I_{dec}$, is trained to reconstruct images $I$. The second one, defined by $f^\mathcal{G}_{enc}$ and $f^\mathcal{G}_{dec}$, is trained using graphs. In both cases, the bottleneck latent representation is modelled by fully connected neurons and have the same size. The encoder/decoder blocks are implemented using standard convolutional layers in the first case, and spectral convolutions in the later case. As depicted in Figure \ref{fig:workflow}, once both models are trained, we decouple the encoders and decoders, keeping only the image encoder $f^I_{enc}$ and graph decoder $f^\mathcal{G}_{dec}$. The HybridGNet is thus constructed by connecting these two networks as $f_{\mathrm{HybridGNet}}(x) = f^\mathcal{G}_{dec}(f^I_{enc}(x))$. The coupled model is initialized with the pre-trained weights, and re-trained until convergence using the landmarks MSE as reconstruction loss, and the KL divergence term for regularization. Thus, HybridGNet takes images $I$ as input and produce graphs $\mathcal{G}$ as output, combining standard with spectral convolutions in a single model.\\

\noindent \textbf{Dual models.}
Based on the HybridGNet model, we implemented two more architectures incorporating a second decoder which outputs dense segmentation masks. The first one is the so called HybridGNet \textit{Dual} model, which has two decoders: the graph $f^\mathcal{G}_{dec}$ and an extra $f^I_{dec}$ whose output is a dense probabilistc segmentation. The model is trained to minimize a reconstruction loss $\mathcal{L}_{dual} = \mathcal{L}_{rec} + \mathcal{L}_{seg}$, combining the landmark localization error ($\mathcal{L}_{rec}$) with the dense segmentation error ($\mathcal{L}_{seg}$). 
The second model, entitled HybridGNet \textit{Dual SC} incorporates skip connections (implemented as sums) between corresponding blocks from the image encoder $f^I_{enc}$ and decoder $f^I_{dec}$. Even though in this paper we do not focus on dense segmentation models, these models were implemented to evaluate the effect of incorporating dense masks into the learning process. Since the dataset only contains contours, we derive the corresponding dense masks for training by filling the organ contours, assigning different labels to every organ.\\

\noindent \textbf{Baseline models.} Our work builds on the hypothesis that encoding connectivity information through graph structures will result in richer representations than standard landmark-based point distribution models. To assess the validity of this hypothesis, we construct standard point distribution models (like those used in ASM) from the graph representations. Differently from the graph structures which incorporate connectivity information, here the landmarks are treated as independent points. For a given graph $\mathcal{G} = <V, \mathbf{A}, \mathbf{X}>$, we construct a vectorized representation $S$ by concatenating the rows of $\mathbf{X}$ in a single vector as $S = (\mathbf{X}_{0,0}, \mathbf{X}_{0,1}, \mathbf{X}_{1,0}, \mathbf{X}_{1,1}, ... \mathbf{X}_{|V|-1,0}, \mathbf{X}_{|V|-1,1})$.

We implement two baseline models using the vectorized representation. In the first one, similar to \cite{milletari2017integrating,bhalodia2018deepssm}, we simply use principal component analysis (PCA) to turn the vectorized representations $S$ into lower-dimensional embeddings. We then optimize the pre-trained image encoder $f_{enc}$ to produce the coefficients of the PCA modes, reconstructing the final landmark-based segmentation as a linear combination of the modes. In the second model, we train a fully connected VAE, $f^S$, to reconstruct the vectorized representations $S$. Following the same methodology used to construct the HybridGNet, we keep only the decoder $f^S_{dec}$ and connect it with the pre-trained image encoder $f^I_{enc}$. These baselines are also trained end-to-end computing the loss on the final location of the reconstructed landmarks, and using the pre-trained weights for initialization.

We also include a third baseline model for comparison, which uses a multi-atlas segmentation (MAS) approach \cite{alven2019shape,alven2016shape}. Given a test image for which we want to predict the landmarks, we take the 5 most similar images (in terms of mutual information) from the training set. We then perform pairwise non-rigid registration (with affine initialization) using Simple Elastix \cite{marstal2016simpleelastix}. The final landmark-based segmentation is obtained by simply averaging the position of the corresponding transferred landmarks to the target space.\\

\noindent \textbf{Detailed architectures and training details.}
The VAE image encoder $f^I_{enc}$ consists of 5 residual blocks \cite{he2015deep}, alternated with 4 max pooling operations. The decoder $f^I_{dec}$ is a mirrored version of $f^I_{enc}$. The graph autoencoder consists on 4 layers of Chebyshev convolutions for the encoder $f^\mathcal{G}_{enc}$ and 5 layers for the decoder $f^\mathcal{G}_{dec}$, with 16 feature maps and a polynomial order of 6. In both models the latent space was fixed to 64 features and implemented as a fully connected layer, see Tables 1-4 in Appendix for more details. The autoencoders were trained using MSE as $\mathcal{L}_{rec}$ (for both the image and landmark reconstruction). The HybridGNet was initalized with the pre-trained weights and refined for 2000 epochs using the MSE to compute the landmark reconstruction error. The dual models also incorporated the $\mathcal{L}_{seg}$ term based on the average between soft Dice \cite{milletari2016v} and Cross Entropy. All models used the KL divergence for regularization (with  $w=1e-5$, choosen by grid search) and were trained using Adam optimizer. All the models were implemented in PyTorch 1.7.1 and PyTorch Geometric 1.6.3. Our code is available at \url{https://github.com/ngaggion/HybridGNet}.

\begin{figure}[t!]
\includegraphics[width=\textwidth]{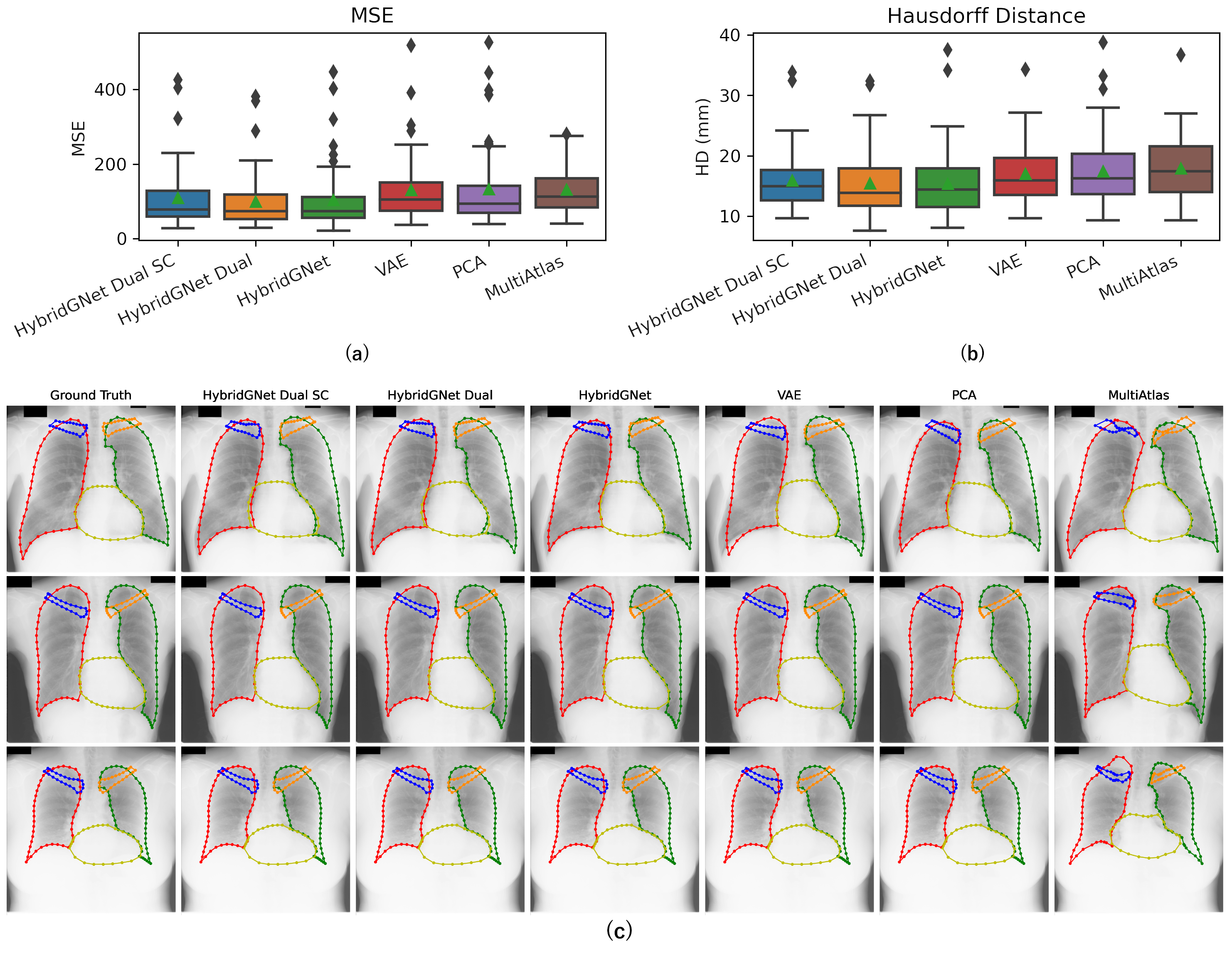}
\caption{Quantitative and qualitative results. The difference between the means of the HybridGNet variants and the baselines for MSE (a) and HD (b) are statistically significant according to a Wilcoxon paired test (see Figure \ref{fig:pvalues} in Appendix for details). We also include qualitative results (c) reflecting the improvement in anatomically plausibility obtained when using the HybridGNets (particularly in the clavicles which are the most challenging).} 
\label{fig:results}

\end{figure}
\section{Experiments and discussion}
We evaluated the proposed approaches using the Japanese Society of Radiological Technology (JSRT) Database \cite{shiraishi2000development}, which consists on 247 high resolution X-Ray images, with expert annotations for lung, hearth and clavicles \cite{gt_van_ginneken:2006-1223}. Annotations consist on 166 landmarks for every image. We constructed the graph adjacency matrix following a connectivity pattern as shown in Figure \ref{fig:results}.c. This matrix was shared across all images. The dataset was divided in 3 folds, using 70\% of the images for training, 10\% for validation and 20\% for testing. We showcase the potential of the proposed HybridGNet architecture in three different experiments.\\



\begin{figure}[t!]
\centering
\includegraphics[width=\textwidth]{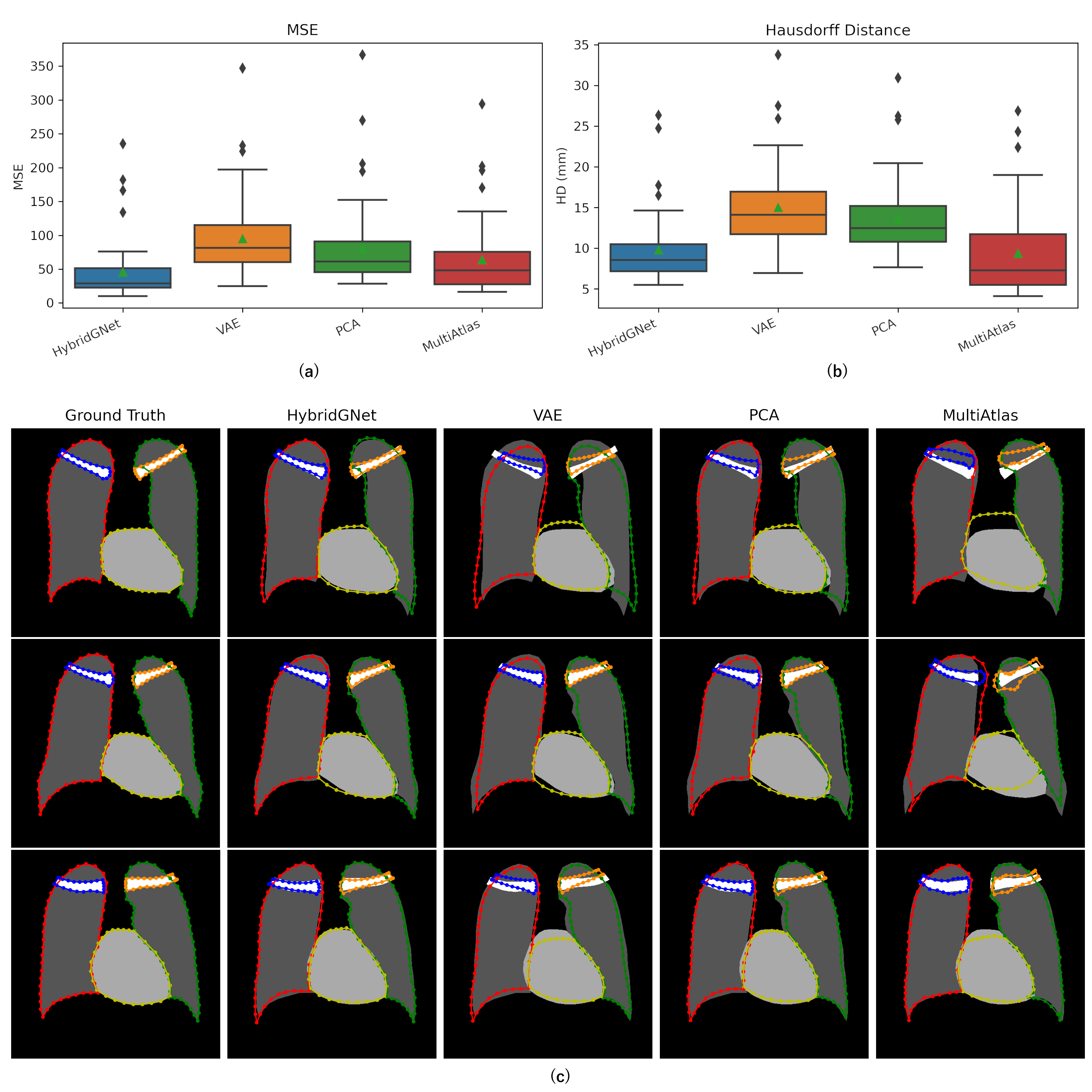}
\caption{Figures (a) and (b) includes quantitative results for Experiment 2, showing that HybridGNet is more accurate at recovering landmark contours from dense segmentation masks. Figure (c) shows examples of landmark-based representations generated from dense segmentation masks in the Experiment 2.} 
\label{fig:seg2}
\end{figure}

\begin{figure}[t!]
\centering
\includegraphics[width=\textwidth]{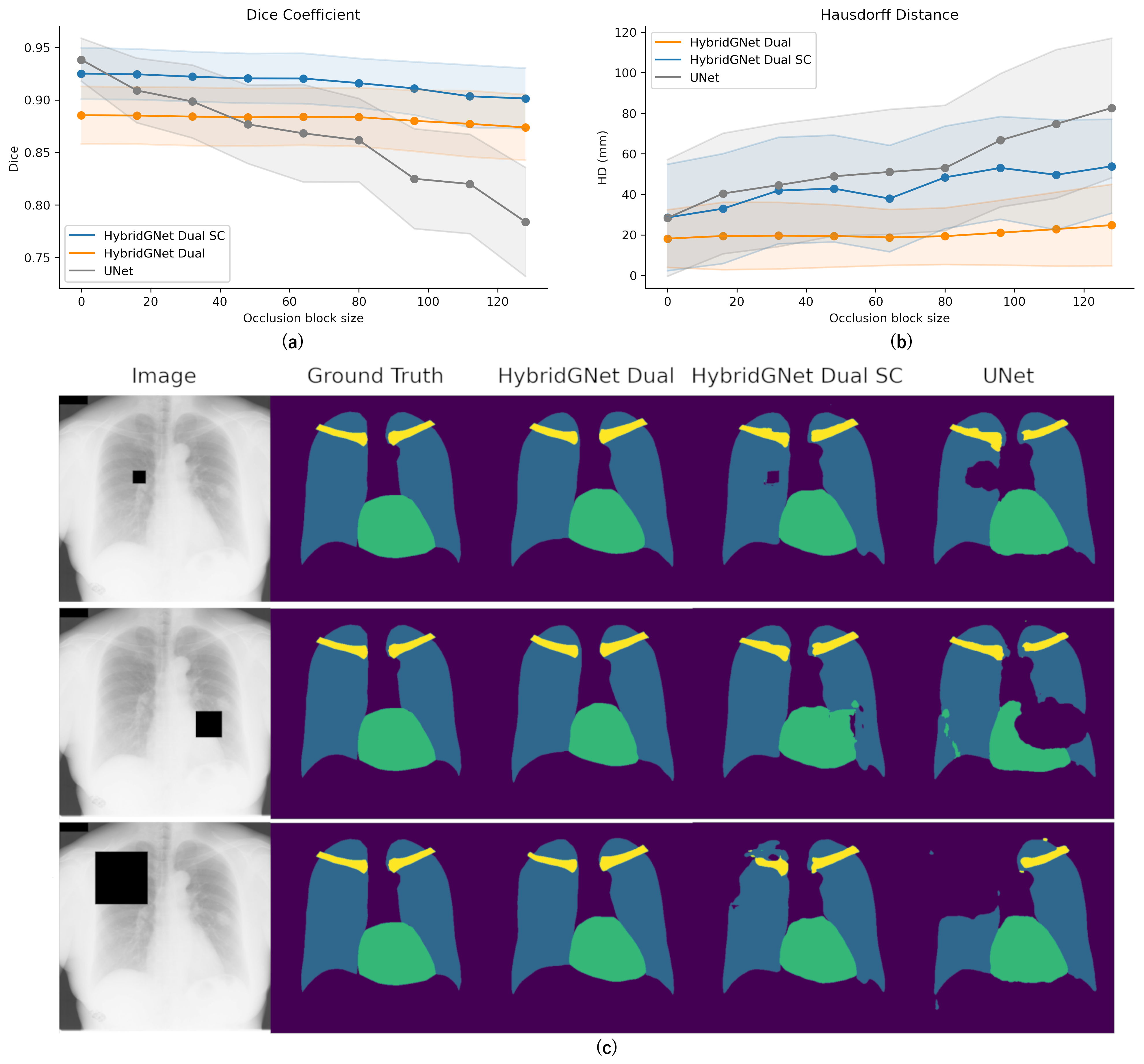}
\caption{Figures (a), (b) and (c) show quantitative and qualitative results for the Experiment 3 (oclussion study). We can see that the HybridGNet Dual models are much more robust to missing parts than a standard UNet model.} 
\label{fig:oclusion}
\end{figure}

\noindent \textbf{Experiment 1: Anatomical landmark-based segmentation.} We evaluated the proposed models using landmark specific metrics, including the mean square error (MSE, in pixel space) over the vectorized landmark location and the contour Hausdorff distance (HD, in millimeters). Figure \ref{fig:results} shows that the Hybrid models outperform the baselines in terms of both MSE and HD. However, incorporating the dual path with dense mask reconstruction does not seem to make a difference.\\

\noindent \textbf{Experiment 2: Generating landmark-based representations from dense segmentations.} The proposed HybridGNet offers a natural way to recover landmark based representations from dense segmentation masks. This could be useful in scenarios where dense segmentation masks are available, but we require matched landmarks for statistical shape analysis. We trained our HybridGNet using dense segmentation masks as input, 
and compared the resulting model with the other approaches trained in the same conditions. Figure \ref{fig:seg2} shows that proposed HybridGNet outperformed the three baselines, confirming that it can be used to construct statistical shape models with landmark correspondences from pixel level annotations.\\

\noindent \textbf{Experiment 3: Robustness study.} We assessed the robustness of the proposed model to image occlusions which may be due to anonymization reasons. We created artificial occlusions by overlapping random black boxes of different size on the test images. In order to highlight the advantages of the HybridGNet over standard dense segmentation models, we trained a UNet architecture (same as the HybridGNet Dual SC but without the graph decoder). Figure \ref{fig:oclusion} shows the results when comparing the dense UNet with our HybridGNet variants. Since we cannot compute the landmark MSE for the dense segmentations, we compared the models evaluating the Dice coefficient and Hausdorff distance on the dense masks obtained from the UNet and the convolutional decoder of the dual models. We can see that not only the UNet, but also the dual models which incorporate dense segmentation masks during training, degrade the performance faster than the HybridGNet as we increase the size of the occlusion block.

\section{Conclusions} In this work we proposed the HybridGNet architecture to perform landmark-based anatomical segmentation. We show that incorporating connectivity information through the graph adjacency matrix helps to improve the accuracy of the results when compared with other landmark-based models which only employ vectorized landmark representations. We showcased different application scenarios for the HybridGNet and confirm that it is robust to image occlussions, in contrast to standard dense segmentation methods which tend to fail in this task.\\

\noindent \textbf{Acknowledgments.} This work was supported by grants from ANPCyT (PICT 2018-3907 and 3384), UNL (CAI+D 50220140100-084LI, 50620190100-145LI and 115LI) and The Royal Society (IES/R2/202165). We gratefully acknowledge the support of NVIDIA Corporation with the donation of the Titan Xp used for this research.

\bibliographystyle{splncs04}
\bibliography{main}

\newpage
\appendix

\section{Appendix}

\begin{table}[]
\centering
\begin{tabular}{|l|c|c|c|}
\hline
\multirow{2}{*}{\textbf{Convolutional Encoder}} & \multicolumn{2}{c|}{Feature Maps} & \multirow{2}{*}{Image Size} \\ \cline{2-3}
                                                & In               & Out            &                             \\ \hline
Residual Block                                  & 1                & 8              & 512x512                     \\ \hline
Max Pooling                                     &                  &                & 256x256                     \\ \hline
Residual Block                                  & 8                & 16             & 256x256                     \\ \hline
Max Pooling                                     &                  &                & 128x128                     \\ \hline
Residual Block                                  & 16               & 32             & 128x128                     \\ \hline
Max Pooling                                     &                  &                & 64x64                       \\ \hline
Residual Block                                  & 32               & 64             & 64x64                       \\ \hline
Max Pooling                                     &                  &                & 32x32                       \\ \hline
Residual Block                                  & 64               & 64             & 32x32                       \\ \hline
Flatten                                         &                  &                &                             \\ \hline
Fully Connected (Mu)                            & 65536            & 64             & -                           \\ \hline
Fully Connected (Sigma)                         & 65536            & 64             & -                           \\ \hline
\end{tabular}
\caption{Convolutional encoder $f^I_{enc}$ detailed architecture.} 
\end{table}

\begin{table}[h]
\centering
\begin{tabular}{|l|c|c|c|}
\hline
\multirow{2}{*}{\textbf{Convolutional Decoder}} & \multicolumn{2}{c|}{Feature Maps} & \multirow{2}{*}{Image Size} \\ \cline{2-3}
                & In                    & Out                   &         \\ \hline
Fully Connected & 64                    & 65536                 & -       \\ \hline
Unflatten       & 65536                 & 64                    & 32x32   \\ \hline
Upsampling      & \multicolumn{1}{l|}{} & \multicolumn{1}{l|}{} & 64x64   \\ \hline
Residual Block  & 64                    & 64                    & 64x64   \\ \hline
Upsampling      & \multicolumn{1}{l|}{} & \multicolumn{1}{l|}{} & 128x128 \\ \hline
Residual Block  & 64                    & 32                    & 128x128 \\ \hline
Upsampling      & \multicolumn{1}{l|}{} & \multicolumn{1}{l|}{} & 256x256 \\ \hline
Residual Block  & 32                    & 16                    & 256x256 \\ \hline
Upsampling      & \multicolumn{1}{l|}{} & \multicolumn{1}{l|}{} & 512x512 \\ \hline
Residual Block  & 16                    & 8                     & 512x512 \\ \hline
Conv2D          & 8                     & 4                     & 512x512 \\ \hline
\end{tabular}
\caption{Convolutional decoder $f^I_{dec}$ detailed architecture.} 
\end{table}

\begin{table}[]
\centering
\begin{tabular}{|l|c|c|c|}
\hline
\multirow{2}{*}{\textbf{Encoder HybridGNet}} & \multicolumn{2}{c|}{Nodes x Nº Kernels} & \multirow{2}{*}{Polynomial Order} \\ \cline{2-3}
                      & Input  & Output &   \\ \hline
Chebyshev Convolution & 166x2  & 166x16 & 6 \\ \hline
Chebyshev Convolution & 166x16 & 166x16 & 6 \\ \hline
Chebyshev Convolution & 166x16 & 166x16 & 6 \\ \hline
Chebyshev Convolution & 166x16 & 166x16 & 6 \\ \hline
Fully Connected       & 166x16 & 64     & - \\ \hline
\end{tabular}
\caption{Graph encoder $f^g_{enc}$ detailed architecture.} 
\end{table}

\begin{table}[]
\centering
\begin{tabular}{|l|c|c|c|}
\hline
\multirow{2}{*}{\textbf{Decoder HybridGNet}} & \multicolumn{2}{c|}{Nodes x Nº Kernels} & \multirow{2}{*}{Polynomial Order} \\ \cline{2-3}
                      & Input  & Output &   \\ \hline
Fully Connected       & 64     & 166x16 & - \\ \hline
Chebyshev Convolution & 166x16 & 166x16 & 6 \\ \hline
Chebyshev Convolution & 166x16 & 166x16 & 6 \\ \hline
Chebyshev Convolution & 166x16 & 166x16 & 6 \\ \hline
Chebyshev Convolution & 166x16 & 166x16 & 6 \\ \hline
Chebyshev Convolution & 166x16 & 166x2  & 6 \\ \hline
\end{tabular}
\caption{Graph decoder $f^g_{dec}$ detailed architecture.} 
\end{table}

\begin{figure}[t!]
\centering
\includegraphics[width=0.9\textwidth]{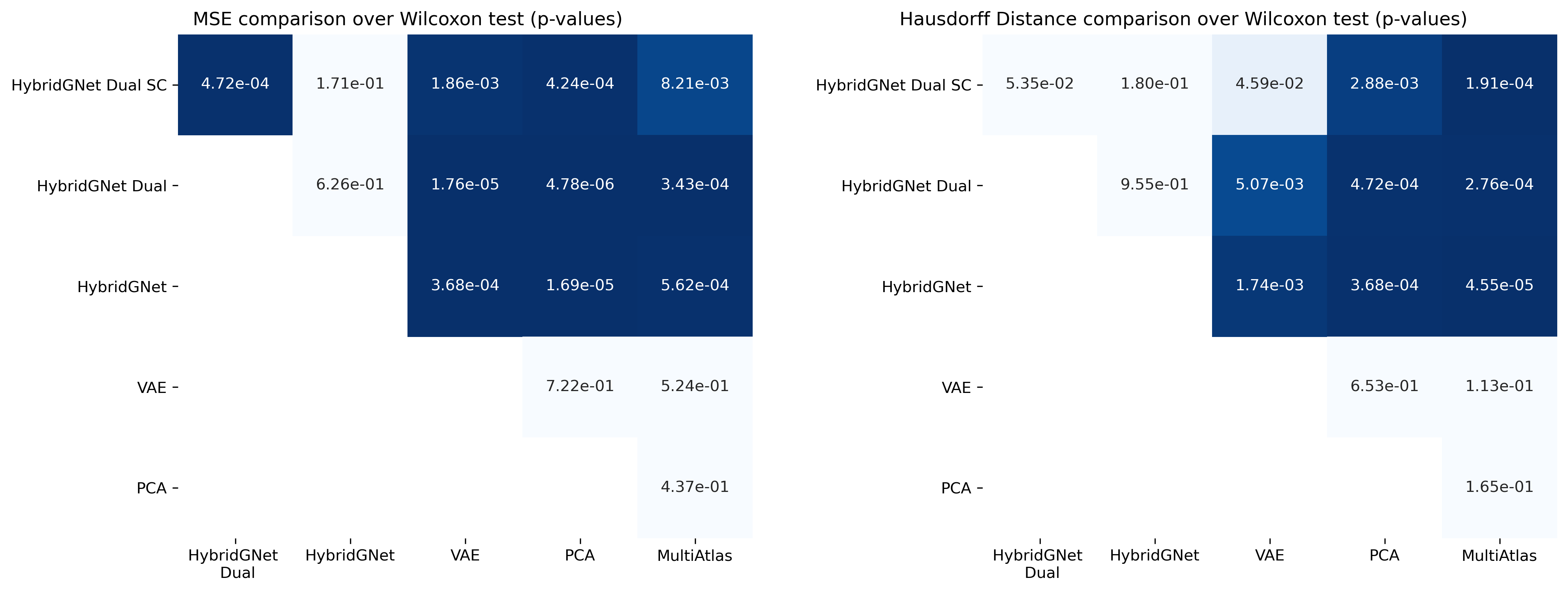}
\caption{Statistical significance for model comparison in Figure 2.a and 2.b of the main manuscript, considering both MSE and Hausdorff Distance via Wilcoxon paired test. P-values $<$ 0.05 implies significative differences between the means.} 
\label{fig:pvalues}
\end{figure}

\end{document}